\documentclass[aps,prb,twocolumn,floatfix,showpacs,groupedaddress]{revtex4}
\usepackage[dvips]{graphicx}

\begin{document}
\title{Local-field effects in current transport through molecular 
electronic devices: Current density profiles and local non-equilibrium 
electron distributions}

\author{Yongqiang Xue $^{*}$ and Mark A. Ratner}
\affiliation{Department of Chemistry and Materials Research Center, 
Northwestern University, Evanston, Illinois 60208, USA}

\begin{abstract}
We analyze non-equilibrium current transport in molecular electronic devices,  
using as an example devices formed by two terphenyl dithiol molecules 
attached to gold electrodes. Using a first-principles based self-consistent 
matrix Green's function method, we show that the spatially resolved current 
density profiles and local electrochemical potential drops provide valuable 
information regarding the local field effect on molecular transport, which 
depend on the internal structure of the molecules and cannot be obtained 
from measuring the current- and conductance-voltage characteristics 
alone.     
\end{abstract}
\pacs{85.65.+h,73.63.-b,73.40.-c}
\date{\today}
\maketitle


\emph{Introduction.}--- Detailed understanding of electron transport 
through single molecules contacted by metallic electrodes is crucial for the 
development of the emerging technology of molecular 
electronics.~\cite{MEReview} Although much attention has been devoted 
to the measured molecular conductance and its dependence 
on metal-molecule interactions,~\cite{MEReview} less is known 
about the non-equilibrium aspects of the phase-coherent molecular 
transport, which provide information not obtainable 
from the conductance measurement.~\cite{Buttiker93,Xue03,Theory}

Several important effects are of interest for non-equilibrium transport: 
(1) A finite electric field associated with the applied bias voltage will 
induce charge pileup (screening) within the molecular junction. 
The local transport field experienced by the tunneling electron is 
the self-consistent screened field; (2) Since the total current is conserved 
throughout the molecular junction for dc-transport, the spatial distribution 
of the current \emph{density} can be highly non-uniform since the molecules 
are intrinsically inhomogeneous; (3) As current flows, the electrons at 
the source and drain contacts have different electrochemical potentials. 
It is not clear if an effective local electrochemical potential (LEP) can 
be defined everywhere within the molecular junction to characterize the 
electron distribution in the non-linear transport regime.  

\begin{figure}
\includegraphics[height=2.5in,width=2.8in]{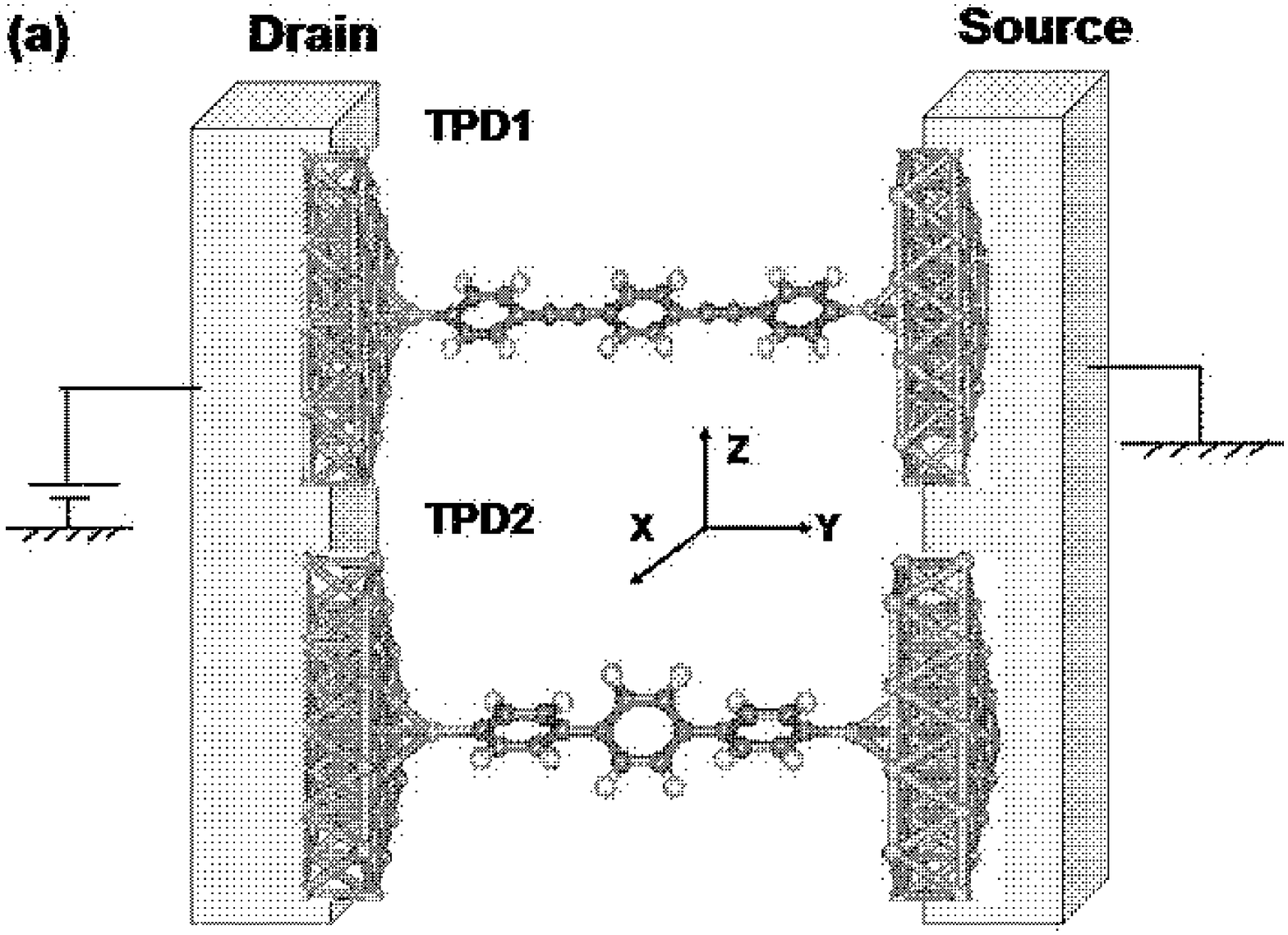} 
\includegraphics[height=2.5in,width=2.8in]{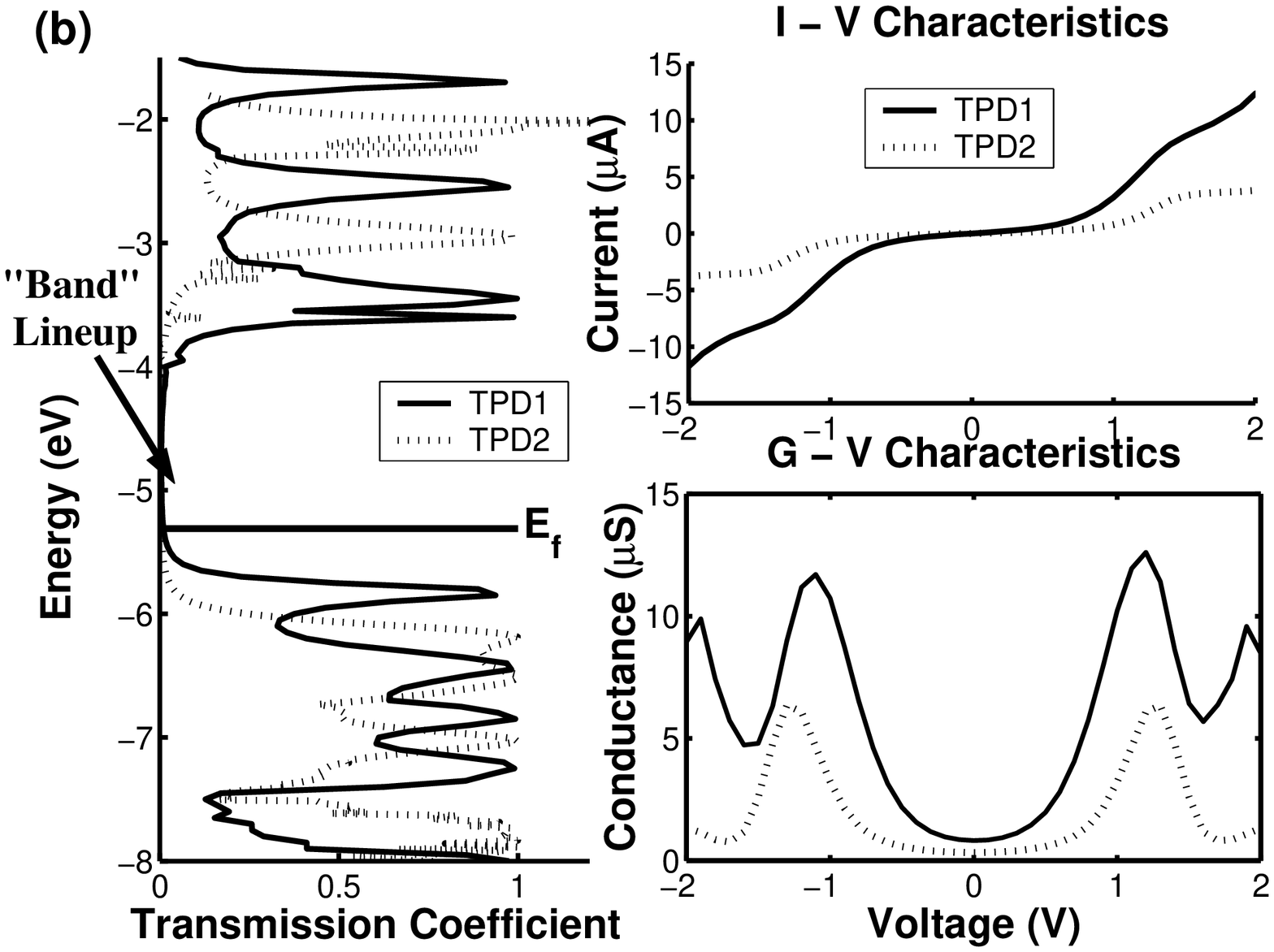} 
\caption{\label{xueFig1} (Color online) 
(a) Schematic illustration of the metal-molecule-metal 
junction for the two terphenyl dithiolate (TPD) molecules. Six gold atoms 
on each metal surface (12 overall) are included into the ``extended 
molecule'' where the self-consistent calculation is performed. The effect of 
the rest of the electrodes (with the 6 atoms on each side \emph{removed}) 
are modeled as self-energy operators. Also shown is the bias polarity and 
coordinate system of the molecular junction; (b) The left figure shows 
the equilibrium electron transmission coefficent. The right 
figures show the self-consistent current-voltage (I-V) and differential 
conductance-voltage (G-V) characteristics of the two molecular junctions. }
\end{figure}

\begin{figure}
\includegraphics[height=2.5in,width=2.8in]{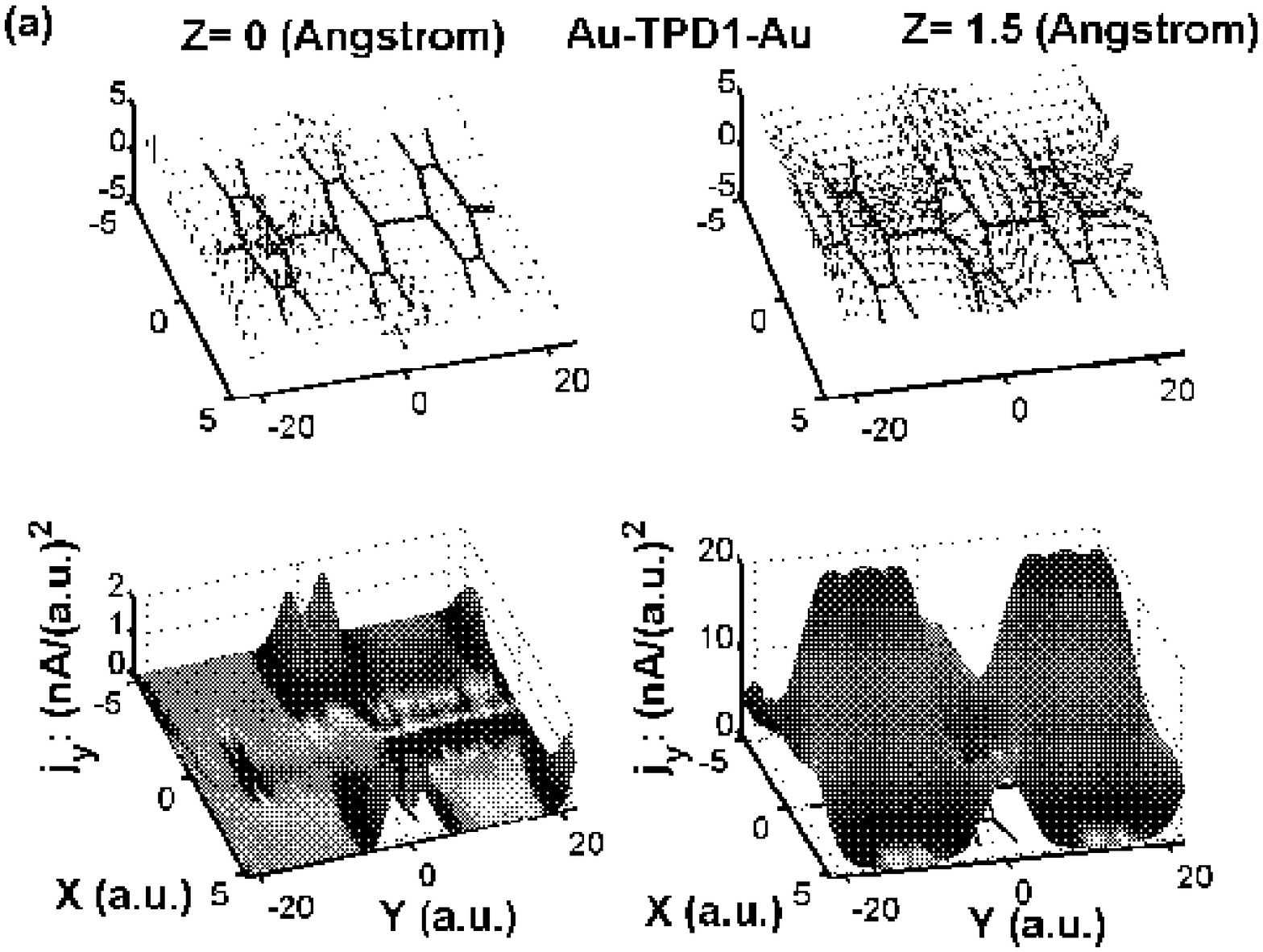}
\includegraphics[height=2.5in,width=2.8in]{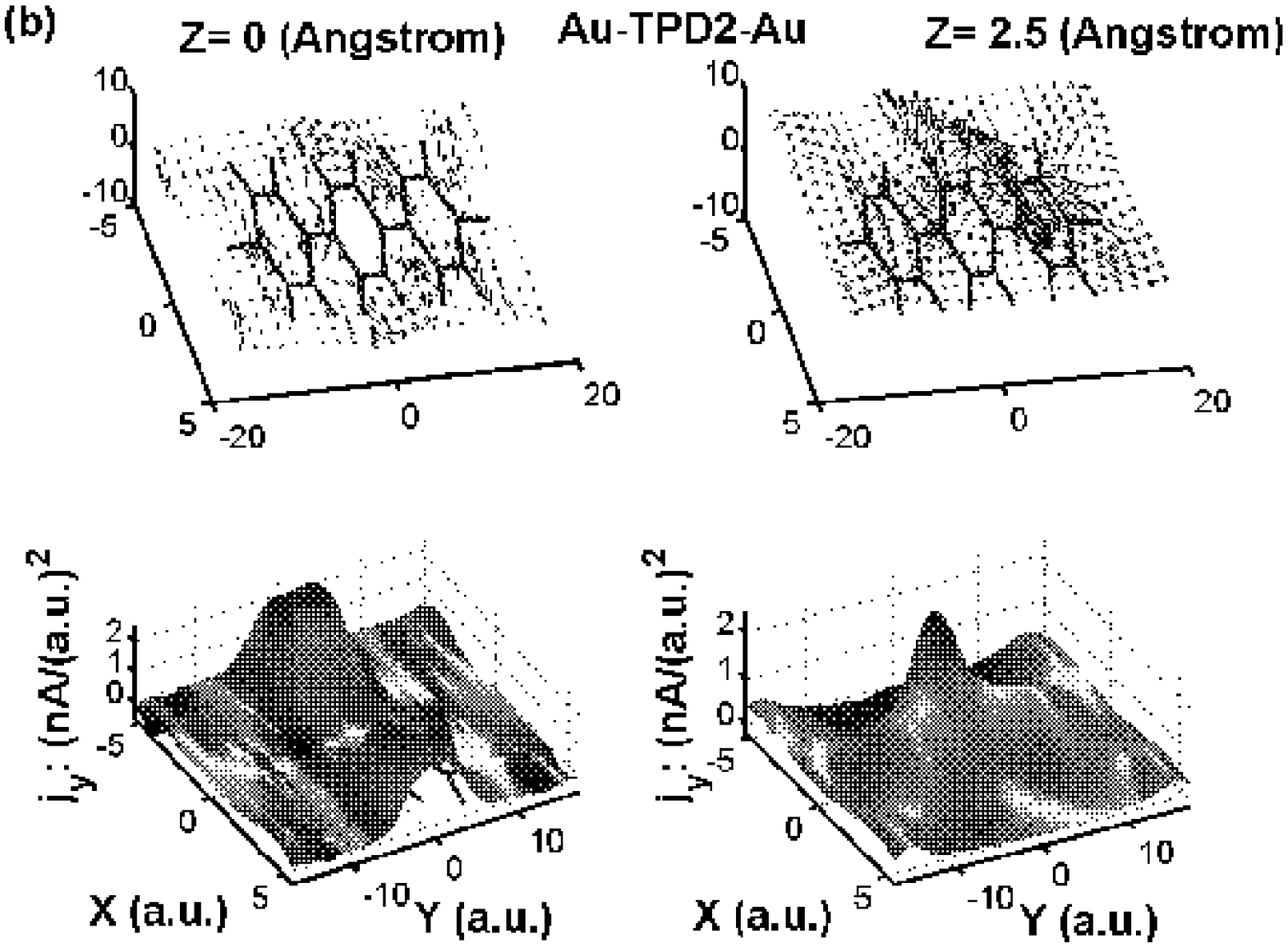}
\caption{\label{xueFig2} (Color online) 
Cross sectional view of current density profiles at 
gold-TPD1-gold (a) and gold-TPD2-gold (b) junctions. Also shown is the 
position of the molecules. For each junction, the upper figures show the 
direction of current density, the lower figures show the magnitude of 
current density component along the transport direction ($Y$ axis here). }
\end{figure}

\begin{figure}
\includegraphics[height=2.5in,width=2.8in]{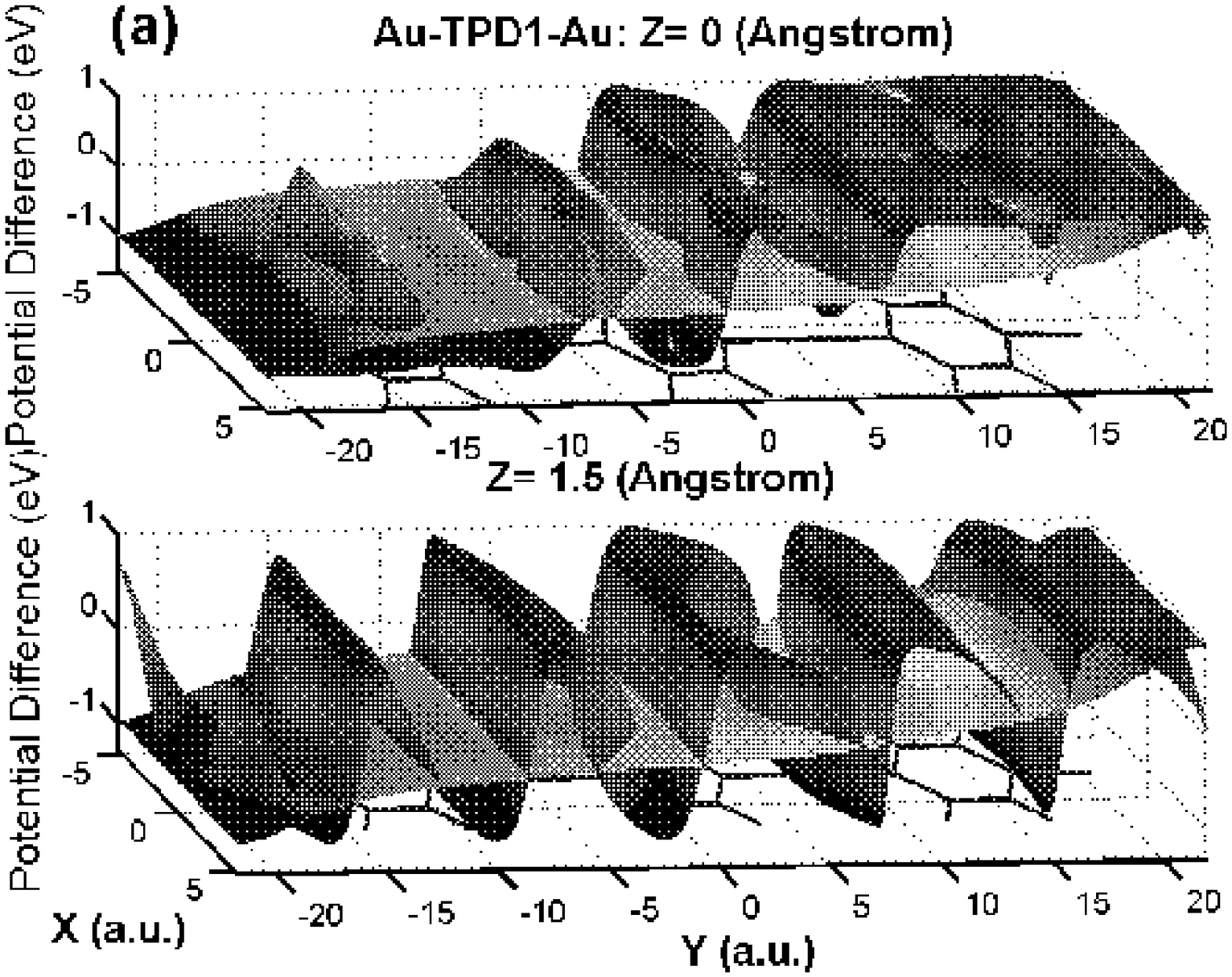}
\includegraphics[height=2.5in,width=2.8in]{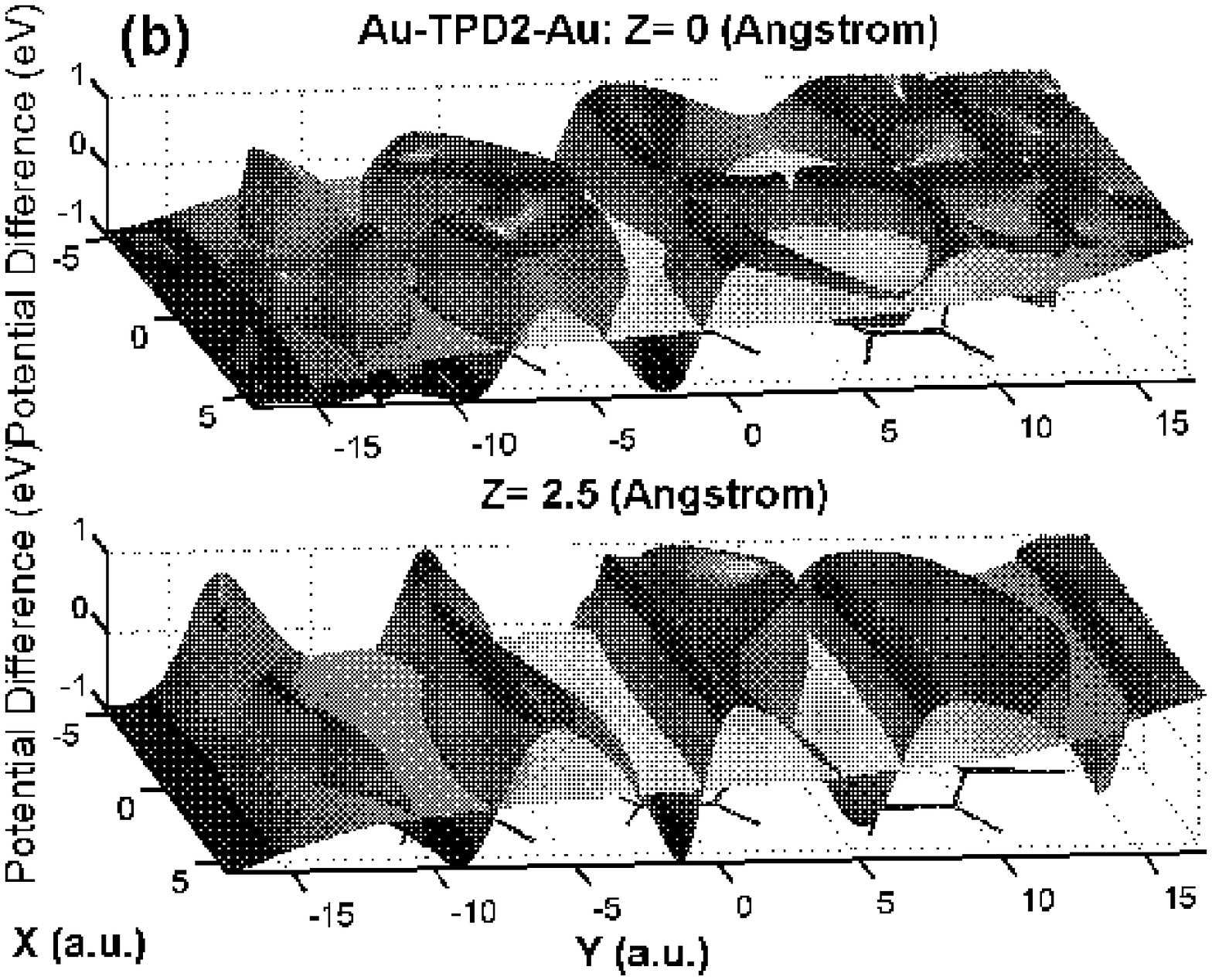}
\caption{\label{xueFig3} (Color online) 
Cross sectional view of local electrostatic potential 
and electrochemical potential drops at the gold-TPD1-gold (a) and 
gold-TPD2-gold (b) junctions. Also shown are the positions of the molecules. 
The electrostatic potential varies rather smoothly from $-V/2$ to $V/2$ 
across the junction, but the electrochemical potential oscillates strongly 
between $-V/2$ and $V/2$.  }
\end{figure}

The screened nature of the local transport field has been analyzed in  
previous works,~\cite{Xue03,Theory} following early ideas of Landauer 
and Buttiker in mesoscopic transport systems.~\cite{LB88} In particular, 
resistivity dipoles and strong local fields have been found in the vicinity 
of potential barriers both at the interface and at certain points 
inside the molecules. These in general modify both the energy and 
wavefunction of the frontier molecular states and lead to different 
electrostatic potential drops depending on the molecules and the device 
structures.~\cite{Xue03} The non-equilibrium electron distribution 
within the molecular junction has also been analyzed 
by projecting the nonequilibrium density-matrix into individual molecular 
orbitals.~\cite{Xue03} Here, we extend previous work by presenting 
real-space analysis of current density profiles and 
local non-equilibrium electron distributions throughout the molecular 
junctions. Such analysis has important implications in resolving the 
following issues: (1) The identification of current path gives a clear 
picture of the conducting and non-conducting part of the molecular 
junction. In particular, large current density may develop at certain 
parts of the molecule, which forms a bottleneck to transport and might 
possibly be imaged using scanned probe microscopes;~\cite{Topinka} 
(2) The local electrochemical potential (LEP) characterizes the 
energy-distribution of the electrons everywhere within the junction, 
and is relevant in multi-probe transport measurement using a weakly 
coupled  phase-sensitive voltage probe,~\cite{Buttiker89} e.g., in 
scanning tunneling potentiometry (STP).~\cite{STP} 
The LEP defined here is determined by the local density 
of states (LDOS) of the molecular junction in the presence of current 
transport, which may possibly be probed using scanning tunneling 
spectroscopy (STS).~\cite{HoMol,Avouris}   

\emph{Non-Equilibrium Green's Function (NEGF) approach to scattering theory 
of molecular transport.}---We analyze the non-equilibrium effects within the 
scattering theory of mesoscopic transport,~\cite{Datta} where the 
relevant physical quantities are obtained from microscopic calculations 
using a self-consistent matrix Green's function theory,~\cite{Xue03,Xue02} 
which combines the NEGF theory~\cite{NEGF} with an effective 
single-particle description of molecular junction electronic structure 
using density functional theory.~\cite{BPW91} We use atomic unit 
throughout the paper unless otherwise noted. 

The central quantity within the NEGF theory is the correlation Green's function 
$G^{<}(\vec r,\vec r';E)=\sum_{ij}G^{<}_{ij}(E) 
\phi_{i}(\vec r)\phi_{j}(\vec r')$, where we have expanded $G^{<}$ in 
terms of local atomic basis sets $\phi_{i}$ (which are real 
functions).~\cite{Xue02} The current density is obtained from 
$G^{<}(\vec r,\vec r';E)$ as~\cite{NEGF} 
\begin{equation} 
\vec j (\vec r)=1/2\lim_{\vec r' \to \vec r} (\nabla^{'}-\nabla)  
\int dE G^{<}(\vec r,\vec r';E)
=1/2\sum_{i,j} (\int dE G^{<}_{i,j}(E) )dS_{ij}(\vec r),
\end{equation} 
where we used the notation $dS_{ij}(\vec r)=
\phi_{i}(\vec r)\nabla \phi_{j}(\vec r)-\phi_{j}(\vec r)\nabla \phi_{i}(\vec r)$.
The terminal current is obtained by integrating the current density over 
a surface separating the molecule from the contact, which leads to the 
standard formula~\cite{Datta,NEGF,Xue02}  
$I_{L(R)}=\int dE Tr \{ \Gamma_{L(R)}[f(E-\mu_{L(R)})A(E)+iG^{<}(E)] \} 
=\int dE \int d\vec r \{ \Gamma_{L(R)}(\vec r,\vec r) \sum_{ij} 
[f(E-\mu_{L(R)})A(\vec r,\vec r;E) +  iG^{<}(\vec r,\vec r;E)] \}$, 
where  $A(\vec r,\vec r;E)=\sum_{ij} A_{ij}(E) \phi_{i}(\vec r)\phi_{j}(\vec r)$. 
The spectral function $A$ and the operators $\Gamma_{L(R)}$ describing 
the contacts are defined in the standard manner.~\cite{Datta,NEGF,Xue02} 
Note that in the matrix notation, we have  
$A=G^{R}[\Gamma_{L}+\Gamma_{R}] G^{A}=A_{L}+A_{R}$ and 
$-iG^{<}=G^{R}[\Gamma_{L}f_{L}+\Gamma_{R}f_{R}]G^{A}
=A_{L}f_{L}+A_{R}f_{R}$, where $A_{L(R)}=G^{R}\Gamma_{L(R)}G^{A}$. 
Defining a local transmission function for electron injection from the 
left (right) electrode 
$T_{L(R)}(\vec r ;E)=\Gamma_{L(R)} (\vec r, \vec r;E)A(\vec r,\vec r;E) $, the 
total current can be expressed in terms of $T_{L(R)}(\vec r ;E)$ and an 
\emph{effective} local electron-distribution function $f_{eff}(\vec r;E)$ as:
$I_{L(R)}=\frac{e}{h} 
\int dE \int d\vec r T_{L(R)}(\vec r;E) [f_{L(R)}(E)-f_{eff}(\vec r;E)]$, 
where $A (\vec r, \vec r;E)f_{eff}(\vec r;E)=-iG^{<}(\vec r,\vec r;E)
=A_{L}(\vec r,\vec r;E)f_{L}(E)+A_{R}(\vec r,\vec r;E)f_{R}(E)$. 

Replacing the spectral functions by the corresponding density of states 
$A(\vec r,\vec r;E)=2\pi \rho(\vec r;E), A_{L(R)}(\vec r,\vec r;E)
=2\pi \rho_{L(R)}(\vec r;E)$, we arrive at the 
following definition of an effective local non-equilibrium electron 
distribution function: 
$f_{eff}(\vec r;E)=\frac{\rho_{L}(\vec r;E)}{\rho(\vec r;E)}f_{L}(E)
+\frac{\rho_{R}(\vec r;E)}{\rho(\vec r;E)}f_{R}(E)$. At low temperature 
and within the linear-response regime, we can replace the Fermi 
distributions by step-like functions and define the local electrochemical 
potential (LEP) as 
$\mu_{eff}(\vec r)=\frac{\rho_{L}(\vec r;E_{f})}{\rho(\vec r;E_{f})}\mu_{L}
+\frac{\rho_{R}(\vec r;E_{f})}{\rho(\vec r;E_{f})}\mu_{R}$ 
where $E_{F}$ is the equilibrium Fermi-level. At finite temperature in the  
non-linear transport regime, the above definition of LEP remains useful for 
understanding transport physics if the coefficients 
$\frac{\rho_{L(R)}(\vec r;E)}{\rho(\vec r;E)}$ vary slowly with 
energy between $\mu_{L}$ and $\mu_{R}$. Since this is true for molecules 
chemisorbed onto the metal surfaces, we get the following LEP defined 
everywhere within the molecular junction: 
\begin{equation}
\mu_{eff}(\vec r)=\frac{\sum_{ij}\rho_{L;ij}(E_{f})\phi_{i}(\vec r)\phi_{j}
(\vec r)}{\sum_{ij} \rho_{ij}(E_{f})\phi_{i}(\vec r)\phi_{j}(\vec r)}\mu_{L}
+\frac{\sum_{ij}\rho_{R;ij}(E_{f})\phi_{i}(\vec r)\phi_{j}(\vec r)}
{\sum_{ij} \rho_{ij}(E_{f})\phi_{i}(\vec r)\phi_{j}(\vec r)}\mu_{R}
\end{equation}
Note that an equivalent definition of the local electrochemical potential 
has been given by Gramespacher and B\"{u}ttiker~\cite{GBSTM} for 
a mesoscopic conductor within the scattering matrix theory. Here 
the contact-resolved local density-of-states 
$\rho_{L(R)}=G^{R}\Gamma_{L(R)}G^{A}/2\pi $ 
is proportional to the injectivity of the left (right) electrode given by 
Gramespacher and B\"{u}ttiker. The LEP $\mu_{eff}(\vec r)$ as defined 
here in a current-carrying conductor may possibly be measured by an ideal 
phase-sensitive non-invasisve voltage probe (B\"{u}ttiker 
probe) introduced at point $\vec r$.~\cite{Buttiker89,GBSTM}  
In scanning tunelling potentiometry, this is identified with the bias voltage 
between the STP tip and the sample under the condition of zero tunneling 
current.~\cite{Buttiker89,STP}  

\emph{Results and their interpretation.}---Since local-field effects are 
most clearly seen for longer molecules with internal barriers,~\cite{Xue03} 
we apply our theory to the devices formed by two extended $\pi$ 
systems--a three-ring oligomer of phenylene ethynylene with dithiol 
substituents (called TPD1) and a terphenyl dithiol molecule (called 
TPD2)--in contact with gold electrodes through the end sulfur atoms. 
The device structure is illustrated schematically in Fig.\ \ref{xueFig1}(a). 
The benzene rings are co-planar when connected by a triple-bonded 
C-C bridge (TPD1), which leads to optimal orbital overlap. In the absence 
of the C-C bridge (TPD2), a torsion angle of $36^{\circ}$ is induced 
between neighbor benzene rings, which leads to weaker orbital overlap 
and an effective potential barrier for electron motion inside the 
molecule.~\cite{Xue03} The calculation is performed using a modified 
version of Gaussian98 program~\cite{G98} with the Becke-Perdew-Wang 
parameterization of density-functional theory~\cite{BPW91} and 
appropriate pseudopotentials with corresponding optimized Gaussian 
basis sets. The details of the self-consistent calculation have been 
discussed extensively elsewhere.~\cite{Xue03,Note} 

The equilibrium (zero bias) electron transmission characteristics of the 
molecular junctions are shown in Fig.\ \ref{xueFig1}(b), from which we find 
that for both molecules, the highest-oocupied-molecular-orbital (HOMO) 
lines up closer to the metal Fermi-level $E_{F}$ than the 
lowest-unoccupied-molecular-orbital (LUMO), similar to previous finding on 
phenyl and biphenyl dithiolate molecules.~\cite{Xue03} The lineup scheme 
is more favorable for the planar TPD1 molecule. The room-temperature 
conductance of the molecular junction is $0.82(\mu S)$ and $0.34(\mu S)$ 
for TPD1 and TPD2 junctions respectively. The self-consistent 
current-voltage (I-V) and differential conductance-voltage (G-V) 
characteristics in the bias range of $-2(V)$ to $2(V)$ are shown in 
Fig. \ref{xueFig1}(b). Note that despite the $\ge 0.3(eV)$ difference in 
the HOMO level position with respect to $E_{F}$ at equilibrium, there is 
only a $0.1(V)$ difference in the bias voltage where the junction reaches 
the first conductance peak ($1.2(V)$ and $1.3(V)$ for the TPD1 and 
TPD2 junction respectively). This is due to the stronger bias-induced 
modification of molecular states (static Stark effect) in the TPD2 molecule, 
since the internal barrier across the neighbor benzene rings leads to 
stronger local-field variations inside the molecule core. This pushes 
the HOMO up relative to the equilibrium Fermi-level and correspondingly 
reduces the voltage needed to move the metal Fermi-levels past 
the HOMO level. 

The overall I-V and G-V characteristics of the two molecules show similar 
behavior, but the spatially-resolved current density distributions are quite 
different due to the different molecular structures, which are shown in 
Fig.\ \ref{xueFig2} for the two molecules at bias voltage of $2(V)$. 
Note that the direction of current is reverse to that of the net electron 
density flux due to the negative electron charge. For the planar TPD1 
molecule, we show both the vectorial plot (direction) of current density 
and the magnitude of the y-component (perpendicular to the electrode 
surface) of current density $j_{y}$ in the $xy$-plane (defined by the 
benzene rings) and in the plane located $1.5(\AA)$ above. For the 
nonplanar TPD2 molecule, this is shown in the $xy$-plane (defined by the 
left-most and right-most benzene rings) and in the plane located 
$2.5(\AA)$ above (the carbon and hydrogen atoms in the central benzene 
ring are located at $Z=\pm 0.7(\AA)$ and $\pm 1.3(\AA)$ respectively).   

For the planar TPD1 molecule, the small y-component of the current density 
in the $xy$ plane peaks around the peripheral hydrogen atoms of the 
central benzene ring. $j_{y}$ increases rapidly (and symmetrically) 
moving above/under the benzene plane, peaks near $0.8 (\AA)$ from it and 
then decreases slowly with further increase of $|Z|$. Note that the peak 
location of $j_{y}$ corresponds to the peak location of the density of the 
$\pi$ electrons in the molecule. The direction of the dominant ($\pi$-electron) 
current flow is such that electrons are injected from the right electrode  
mostly along the sulfur-surface bond, and then propagate through the 
middle of the right-most benzene ring (centered around $Y$-axis), being 
diverted to the perimeters of the central benzene, and converge again 
into the middle of the left-most benzene ring. For the nonplanar TPD2 
molecule, there is a large $j_{y}$ in the plane $Z=0$, which resides mostly 
around the central benzene ring. Moving away from $Z=0$, $j_{y}$ 
increases gradually in both the left and right benzene rings, and reaches 
its peak at about $0.7(\AA)$, corresponding to the position of the carbon 
atoms in the central benzene (the peak value is only about three times 
that in the $XY$ plane) and then decreases slowly. For positive 
(negative) $Z$, $j_{y}$ is larger on the negative (positive) $X$ half of the 
junction, determined by the location of carbon and hydrogen atoms in the 
central benzene ring. The direction of the dominant current flow is such 
that electrons propagate rather uniformly through the right-most 
benzene ring, being diverted towards the negative (positive) $X$ half of 
the molecule at positive (negative) $Z$ and propagate again rather 
uniformly through the left benzene ring. 

The local electrostatic potential and electrochemical (LEP) potential 
drops in the molecular junction are shown in Fig.\ \ref{xueFig3} for bias 
voltage of $2(V)$. Unlike the electrostatic potential drop which varies rather 
smoothly from $-V/2$ to $V/2$ across the junction, the local electrochemical 
potential shows oscillatory behavior between $-V/2$ and $V/2$ throughout 
the molecular junction. This is due to the phase-sensitive nature of the 
voltage probe: an electron wave incident from either electrode gives 
two contributions to the electron flux in the voltage probe at 
point $\vec r$, i.e., the direct transmission and the transmission after 
multiple scattering within the molecular junction. The oscillatory behavior 
in LEP therefore demonstrates the presence of a potential barrier for 
electron injection into the molecular junction. The more opaque (smaller 
transmission coefficient) the barrier is, the larger the magnitude of the 
oscillation will be.~\cite{Buttiker89}  
The oscillation of the LEP is also sensitive to the shape of the potential 
barriers, which shows up clearly in the two planes of the molecular 
junctions (Fig. \ref{xueFig3}). The LEP in the TPD2 junction oscillates 
more strongly than the TPD1 junction due to the presence of barriers both 
at the metal-molecule interface and inside the molecule induced by the 
weaker orbital overlap across neighbor benzene rings. In addition, there is 
also oscillation in the direction (along $X$ axis) perpendicular to the 
transport direction due to the three-dimensional structure of the molecular 
junction. 

\emph{Discussion and Conclusion.} Although in principle both the local 
electrochemical potential (LEP) and the current density profile in 
single-molecule devices may possibly be measured using scanning 
nanoprobe techniques, in practice this can be extremely difficult due to the 
spatial resolution involved and the requirement that the local probe should 
be minimally invasive. For example, the LEP varies on the scale of the 
Fermi wavelength ($\lambda_{F}$) of electrons injected from the electrodes, 
which  is $\sim 1 (\AA)$ for metallic electrodes, unlike the 
two-dimensional-electron-gas fabricated from semiconductor 
heterostructures where $\lambda_{F}$ can be of 100(nm) 
and longer.~\cite{McEuen} Further work is therefore needed for a 
quantitative evaluation taking into account realistic experimental 
conditions in such multi-probe transport measurements, in particular 
the probe geometry effect and local probe-induced perturbation of 
the junction electronic processes. The present work hightlights the 
importance of local atomic-scale analysis in revealing subtle effects in 
single-molecule electronic devices. 

This work was supported by the DARPA Moletronics program, the DoD-DURINT 
program and the NSF Nanotechnology Initiative. 

\end{document}